\documentclass[11pt]{article}
\usepackage{cite}
\usepackage{bm}
\usepackage{amsmath}
\usepackage{mathtools}
\usepackage{amssymb}
\usepackage{amsthm}
\usepackage{graphicx}
\usepackage{multicol}
\usepackage{dcolumn}
\usepackage{hyperref}
\usepackage{xcolor}
\usepackage{units}
\usepackage{float}
\usepackage{authblk}
\usepackage[shortlabels]{enumitem}

\usepackage[margin=0.5in]{geometry}

\begin{document}

\title{Nonlocal mechanisms of attosecond interferometry in three-dimensional systems}
\author[1]{Denis Jelovina}
\author[2]{Armin Scrinzi}
\author[3]{Hans Jakob W\"orner}
\author[4]{Axel Schild}
\affil[1,3,4]{Laboratorium f\"ur Physikalische Chemie, ETH Z\"urich, Z\"urich, Switzerland}
\affil[2]{Physics Department, Ludwig Maximilians Universit\"at, D-80333 Munich, Germany}

\date{\today}
\maketitle

\begin{abstract}
  Attosecond interferometry (AI) is an experimental technique based on ionizing a system with an attosecond pulse train in the presence of an assisting laser.
  This assisting laser provides multiple pathways for the photoelectron wave packet to reach the same final state, and interference of these pathways can be used to probe properties of matter.
  The mechanism of AI is well-understood for isolated atoms and molecules in the gas phase, but not so much in the condensed phase, especially if the substrate under study is transparent.
  Then, additional pathways open up for the electron due to scattering from neighbouring atoms.
  We investigate to what extent these additional pathways influence the measured photoionization delay with the help of one- and three-dimensional model systems.
  In both cases, we find that the total delay can be expressed as the sum of a local (photoionization) delay and a non-local delay which contains the effect of electron scattering during transport.
  The 1D system shows that the non-local delay is an oscillatory function of the distance between the sites where ionization and scattering take place.
  A similar result is obtained in 3D, but the modulation depth of the non-local delay is found to strongly depend on the effective scattering cross section.
  We conclude that attosecond interferometry of disordered systems like liquids at low photon energies (20-30 eV) is mainly sensitive to the local delay,
  i.e., to changes of the photoionization dynamics induced by the immediate environment of the ionized entity, and less to electron scattering during transport through the medium.
  This conclusion also agrees with the interpretation of recent experimental results.
  
\end{abstract}


\section{\label{sec:Introduction}Introduction}

Attosecond interferometry (AI) 
is a technique to probe matter via ionization.
This technique has been applied to atoms and molecules \cite{klunder2011,guenot2012,guenot2014,palatchi2014,heuser2016,huppert2016,jordan2017}
as well as solids \cite{tao2016,siek2017angular}.
Recently, it has also been extended to liquids \cite{jordan2015photoelectron,IJordanSci2020}.
An attosecond pulse train induces the ionization process and interaction with an assisting laser pulse provides multiple pathways for the ionized system to reach the same final state(s), which leads to interference in the measured photoelectron spectrum.
From this spectrum, information can be gained about the state that the electron was in before ionization, its dynamics following ionization,  and hence about the quantum-mechanical structure of the system under study.

The mechanism underlying AI is well-understood for atoms and molecules in the gas phase.\cite{kheifets2010,nagele2011,pazourek2012,dahlstrom2012,ivanov2012,dahlstrom2013,kheifets2013,maquet2014,huppert2016,hockett2016,baykusheva2017a,baykusheva2017b} 
By means of high-harmonic generation, attosecond extreme ultraviolet (XUV) pulse trains are generated from an assisting femtosecond laser with frequency $\omega$, typically in the infrared (IR) regime.
The XUV pulse train contains only frequencies that are odd multiples of $\omega$, i.e., frequencies $(q + 1) \omega$ for $q \in 2 \mathbb{N}$.
The sample is irradiated with the XUV pulse train and with the assisting femtosecond IR laser that was used to generate the XUV pulse.
Pulses have a well-defined time offset $\Delta t$ with respect to each other, which we define in terms of the relative phase-delay of the IR.
The intensity of the assisting laser is chosen such that the system can absorb or emit an additional IR photon during ionization, thus creating photoelectrons with a kinetic energy $q \omega - E_{\rm b}$ (we use atomic units throughout the article) corresponding to even multiples of $\omega$, where $E_{\rm b}$ is the binding energy of the electron.
The intensity $I_q^{\rm SB}$ of the photoelectron current for a particular sideband $q$ oscillates with $\Delta t$,
\begin{align}
  I_q^{\rm SB} \propto 1 + \cos(2 \omega (\Delta t-\tau)),
  \label{eq:sideband}
\end{align}
due to interference of mainly two pathways: 
Absorption of a photon with frequency $(q+1) \omega$ from the ionizing XUV pulse train and emission of a photon of frequency $\omega$ from the assisting laser, or absorption of a photon with frequency $(q-1) \omega$ and absorption of a photon of frequency $\omega$.
Here, $\Delta t = 0$ is defined as the moment when the maximum of both the XUV pulse and the assisting IR pulse coincide.
The shift $\tau$ of the oscillation relative to this time zero is called the ``ionization delay'' and it is the main experimental observable.

AI experiments or related experiments using attosecond streaking\cite{itatani2002,kienberger2004} have been performed for metals,\cite{cavalieri2007,neppl2012,neppl2015,locher2015,lucchini2015,tao2016} where it was assumed that the assisting laser has a negligible effect inside the medium.
However, liquids as well as dielectrics and wide-bandgap semiconductors are transparent to IR light.
For example, the closely-related technique of attosecond streaking has recently been performed on SiO$_2$ nanoparticles \cite{seiffert2017}.
In a situation where both the ionizing XUV pulse and the assisting IR pulse can enter the medium in the condensed phase, at least two additional effects can be expected compared to the gas phase:
First, the electronic structure of a molecule can be different from the gas phase due to interactions with neighboring molecules in the vicinity of the ionized site.
As the measured delays directly reflect the electronic structure, a difference in the delays of gas phase and condensed phase can yield valuable information about the molecular properties of the latter.
We call the contribution to the ionization delay due to interference of pathways at the ionization site the ``local'' delay $\tau_{\rm l}$.
A second effect is due to the scattering of the ionized electron during its transport.
During each scattering event, photons of the assisting laser can be absorbed or emitted by the system, thus opening up many additional interfering pathways. 
Consequently, from the AI signal also valuable information about the scattering properties of the considered material might be obtained.
This contribution to the ionization delay is the ``non-local'' delay $\tau_{\rm nl}$.
The main problem with analysis of the experimental AI signal for the condensed phase is thus to discern the two effects in order to correctly interpret the results.

The non-local delay due to collisions at places remote from the ionization site has previously been investigated in a one-dimensional model \cite{PhysRevA.97.063415}, where it was found that the total delay is a sum of local and non-local delay,
\begin{align}
  \tau = \tau_{\rm l} + \tau_{\rm nl}.
  \label{eq:tau}
\end{align}
Thus, the two delays can be studied independently.
With the 1D-model it was shown that the non-local delay due to one collision oscillates with the distance to the scattering site, and multiple as well as randomly distributed collisions were also analyzed.
Here, we extend this study by considering the 1D model again and by investigating the manifestations of the effect in three spatial dimensions.
For this purpose, in section \ref{sec:Theory} we present the numerical methods and models used in the study, and we discuss a trajectory-based picture that helps to understand the results.
Thereafter, our results for the non-local delays of the models are presented in section \ref{sec:results}.
Finally, in section \ref{sec:conclusion} we discuss the implications of our study for the analysis of experimental data, in particular regarding AI measurements in the liquid phase.

\section{\label{sec:Theory}Theory}

\subsection{\label{subsec:1D_calculations}1D calculations}
We consider the 1D model from \cite{PhysRevA.97.063415}, where an electron is initially in the ground state of the potential
\begin{align}
  V(x)=-V_1 \frac{e^{-\frac{\left|x\right|}{\lambda}}}{\sqrt{x^2+s^2}}.
\end{align}
The parameters \unit[$V_1=1$]{$E_{\rm h}$}, \unit[$s=1.2741$]{$a_0$}, and \unit[$\lambda=10$]{$a_0$} are chosen such that the ground-state binding energy of this potential is that of a hydrogen atom, i.e., \unit[$E_{\rm b} = - E_0 = 0.5$]{$E_{\rm h}$}, 
an essentially arbitrary choice which does not affect the qualitative results.

During a collision of an electron with an atom or molecule the system can absorb or emit photons, thereby changing the kinetic energy of the electron.
Due to energy and momentum conservation
the photoelectron propagating in a medium, exchanges an additional photon with the laser field only in the presence of a perturbing potential,
where it is not the electron but the scattering complex that is responsible for the actual photon absorption/emission.
For the 1D model we use a Yukawa potential $V_{\rm p}$ to represent such a scattering site centered at $r_{\rm sc}$,
\begin{equation}
  V_{\rm p}(x,r_{\rm sc}) = V_0 \frac{e^{-\frac{\left|x-r_{\rm sc}\right|}{\lambda}}}{\sqrt{\left(x-r_{\rm sc}\right)^2+s^2}},
  \label{eq:yukawa_1D}
\end{equation}
with the height of the potential being $V_0=$\unit[$0.073$]{$E_{\rm h}$}\unit[$=2.0$]{eV},
a choice where the photoelectron energy is higher than the potential barrier height, therefore excluding quantum tunneling effects.

Under influence of the laser field the electron is ejected by XUV photon absorption, after which it can  absorb or emit additional IR photons. 
The electric field $E(t)$ describing electromagnetic radiation is composed of an IR pulse with a duration of \unit[40]{fs} centered at a
wavelength of \unit[800]{nm} (corresponding to the fundamental photon energy \unit[$\hbar\omega=1.55$]{eV}) with intensity \unit[$3.5\cdot 10^{10}$]{W/cm$^2$} and its XUV harmonics $q=15$ and $q=17$
(with frequencies $q \omega$), both with duration of \unit[30]{fs} with intensity \unit[$3.5\cdot 10^8$]{W/cm$^2$}.
In our previous work on the 1D model, we have shown that the XUV pulse duration has no influence on the non-local delay.
Therefore, for computational convenience, we choose relatively long pulses in order to suppress  spectral overlap between sidebands,
which allows a straightforward analysis of the calculated photoelectron spectra.
The XUV pulses are chosen to have constant phase, while the carrier envelope phase (CEP) 
\begin{align}
  \delta = \omega \Delta t
  \label{eq:delta}
\end{align}
 of the IR pulse is varied, which corresponds to changing the time offset $\Delta t$ between the IR pulse and the XUV pulse train. 
We solve the time-dependent Schr{\"o}dinger equation (TDSE) in the dipole approximation and in the length gauge,
\begin{equation}
  i\frac{\partial }{\partial t}  \Psi (x,t) = \left( -\frac{1}{2}\frac{d^2}{{dx}^2} + V(x) + V_p(x) + x E(t) \right)  \Psi (x,t),
\end{equation}
for the 1D system and calculate the photoelectron spectra.
These spectra show the characteristic energy structure with the sideband intensity oscillating as a function of $\Delta t$ as given in Eq. \eqref{eq:sideband}.

\subsection{\label{subsec:3D_calculations}3D calculations}

As a model system in three spatial dimensions, we consider a hydrogen atom placed at the origin of the coordinate system and initially in the ground state with binding energy \unit[$E_{\rm b} = -E_0 = 0.5$]{$E_{\rm h}$}.
To test the effect of a scattering center on the attosecond interference signal, we place an additional repulsive potential $V_{\rm p}(r,\theta)$ (where the sign of the potential changes only the sign of non-local delay)
at some distance from the origin, where $r$ is the radial coordinate and $\theta$ is the polar angle in spherical coordinates. 
The actual potentials are defined below, but we note that all $V_{\rm p}(r,\theta)$ used in our study are cylindrically symmetric around the $\hat{\bm{z}}$-axis.
The system is irradiated by an electric field that is linearly polarized in $\hat{\bm{z}}$-direction. 

\begin{figure}[!h]
  \includegraphics[width=0.9\textwidth]{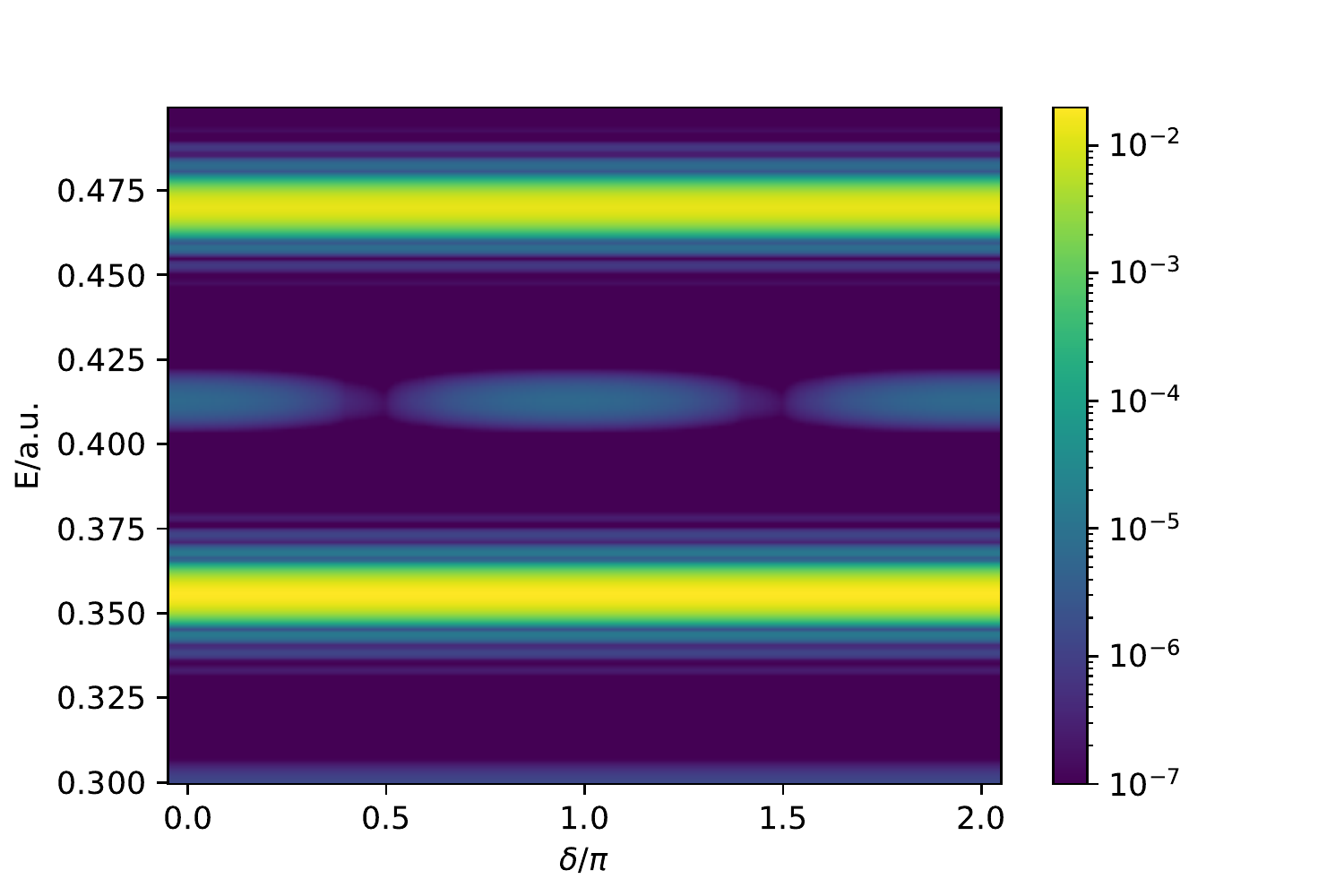}
  \caption{
    Example of an attosecond interference spectrogram for the three-dimensional model of the hydrogen atom.
    The detection angle is aligned with the light polarization ($\theta=0$) direction.
    Clearly visible are the strong main bands and the sideband located at an electron kinetic energy $E$ in between the two main bands.
    The intensity of the sideband oscillates with the phase $\delta$ between the ionizing XUV pulse and the assisting IR pulse.
    }
  \label{fig:rtrace01}
\end{figure}

The wave function for the hydrogen atom is expanded as $\Psi(r,\theta,\phi)=\sum_{l,m,i} a_{l,m,i} f_{i}\left(r\right) Y_l^m(\theta,\phi)$,
where $Y_l^m$ are spherical harmonic functions and $f_i$ are radial functions.
Since the system is cylindrically symmetric around the $\hat{\bm{z}}$-axis, the expansion is limited to $m=0$ terms.
The ground state of the system is calculated by diagonalizing the field-free Hamiltonian of the hydrogen atom $H_0$.
Scattering at the perturber is described by the $V_{\rm p}$ potential and the interaction with the electromagnetic field, $V_{\rm F}(t)$, is treated within the dipole approximation, both in the length gauge $V_{\rm F}(t)=\bm{r}\cdot \bm{F}(t)$ and in the velocity gauge  $V_{\rm F}(t)= \bm{A}(t) \cdot \bm{p}$.
We solve the TDSE
\begin{equation}
  i\frac{\partial }{\partial t}  \Psi (\bm r,t) = \left( H_0 + V_{\rm p} + V_{\rm F}(t) \right)  \Psi (\bm r,t)
\end{equation}
with two methods:
First, we use a finite element discrete variable representation (FEM-DVR) radial basis and exterior complex scaling \cite{McCurdy_2004,PhysRevA.62.032706} for the long-range Coulomb potential, where we extract the photoionization spectrum from the wave function $\Psi(t_{\rm fin})$ at the end of the laser pulse by the surface integral technique \cite{PhysRevA.76.043420,McCurdy_2004}.
Second, we use the tRecX code \cite{scrinzi2012} for short-range binding potentials constructed from Coulomb potentials truncated at \unit[20]{$a_0$} radial distance, with the photoionization spectrum obtained with the t-SURFF method.\cite{scrinzi2012}
To describe the potentials accurately and efficiently, we use a two-center expansion with one center being the binding potential and one being the remote scattering potential.
Comparison of the two methods suggests that the long range interaction is changing only the photoionization time delay but insignificantly affects the effect introduced by scattering.

In the 3D simulations we use the same laser field parameters as for the 1D model.
Like for the 1D simulations, the photoelectron spectra show the characteristic structures of main bands and sidebands, as exemplified in Figure \ref{fig:rtrace01}:
Two main bands are visible at photoelectron kinetic energies $E = q \omega - E_{\rm b}$ corresponding to one-XUV-photon absorption for harmonics $q=15$ and $q=17$, as well as a sideband centered in between bands at \unit[$E=0.41$]{$E_{\rm h}$}, oscillating in intensity with $\delta$ or $\Delta t$ and originating from additional IR-photon absorption, see \eqref{eq:sideband} and \eqref{eq:delta}.

\subsection{\label{subsec:nonlocal_paths}The trajectory-picture and local vs.\ nonlocal pathways}

AI can, to a large extent, be understood in a trajectory picture based on the soft-photon approximation \cite{kroll1973,faisal1973,madsen2005,galan2013}.
This approximation can be used to express the laser-assisted photoionization and scattering processes with transition amplitudes, while the electron propagation between ionization and scattering sites is assumed to be that of a free electron. 
The process of absorbing an XUV photon and then absorbing or emitting another photon with frequency $\omega$ from an assisting laser field is often termed the laser-assisted
photoelectric effect (LAPE). 
Within the soft-photon approximation, the transition amplitudes for a LAPE process can be written as
\begin{equation}
  F^{q+\nu}_{q}=e^{i\nu(\frac{\pi}{2}+\delta)} J_{\nu}\left( \frac{ \bm{k}_q \cdot \bm{F}_0^{\rm IR}}{\omega^2} \right) f^{\rm PI}_{q+\nu},
  \label{eq:lape}
\end{equation}
where an XUV photon of frequency $(q+\nu) \omega$ is absorbed and $\nu$ photons of frequency $\omega$ are absorbed or emitted.
In our case, we have $\nu \in \{-1, 0, 1\}$ because the laser parameters are chosen such that only one-photon exchanges with the assisting laser field happen.
Here and in the following, we write that the electron is in a (continuum) state $q$ if its kinetic energy corresponds to $q \omega - E_{\rm b}$.
Thus, $F^{q+\nu}_{q}$ describes the transition from state $q+\nu$ to state $q$.
The relevant quantities in \eqref{eq:lape} are the field-free photoionization amplitude $f^{\rm PI}_{q+\nu}$ for state $q+\nu$, the momentum $\bm{k}_q$ of the electron in state $q$,
and the maximum amplitude of the assisting laser field $\bm{F}_0^{\rm IR}$.
The functions $J_{\nu}$ are Bessel functions of the first kind.

The soft-photon approximation provides a similar equation for laser-assisted electron scattering (LAES) for a transition from state $q+\nu$ to state $q$ during a collision,
\begin{equation}
  f^{q+\nu}_{q}=e^{i\nu(\frac{\pi}{2}+\delta)} J_{\nu}\left( \frac{(\bm{k}_q - \bm{k}_{q+\nu}) \cdot  \bm{F}_0^{\rm IR}}{\omega^2} \right) f^{\rm ES}_{\bm{k}_q - \bm{k}_{q+\nu}},
  \label{eq:laes}
\end{equation}
where $f^{\rm ES}_{\bm{k}_q - \bm{k}_{q+\nu}}$ is the field-free scattering amplitude
for elastic scattering with momentum transfer $\bm{k}_q - \bm{k}_{q+\nu}$.
With these transition amplitudes, complex-valued electron trajectories (in the sense of products of transition amplitudes that represent the individual pathways which the photoelectron can take)
can be constructed and the observed interference effects in the photoelectron signal can be explained.

\begin{figure}[!h]
  \includegraphics[width=0.9\textwidth]{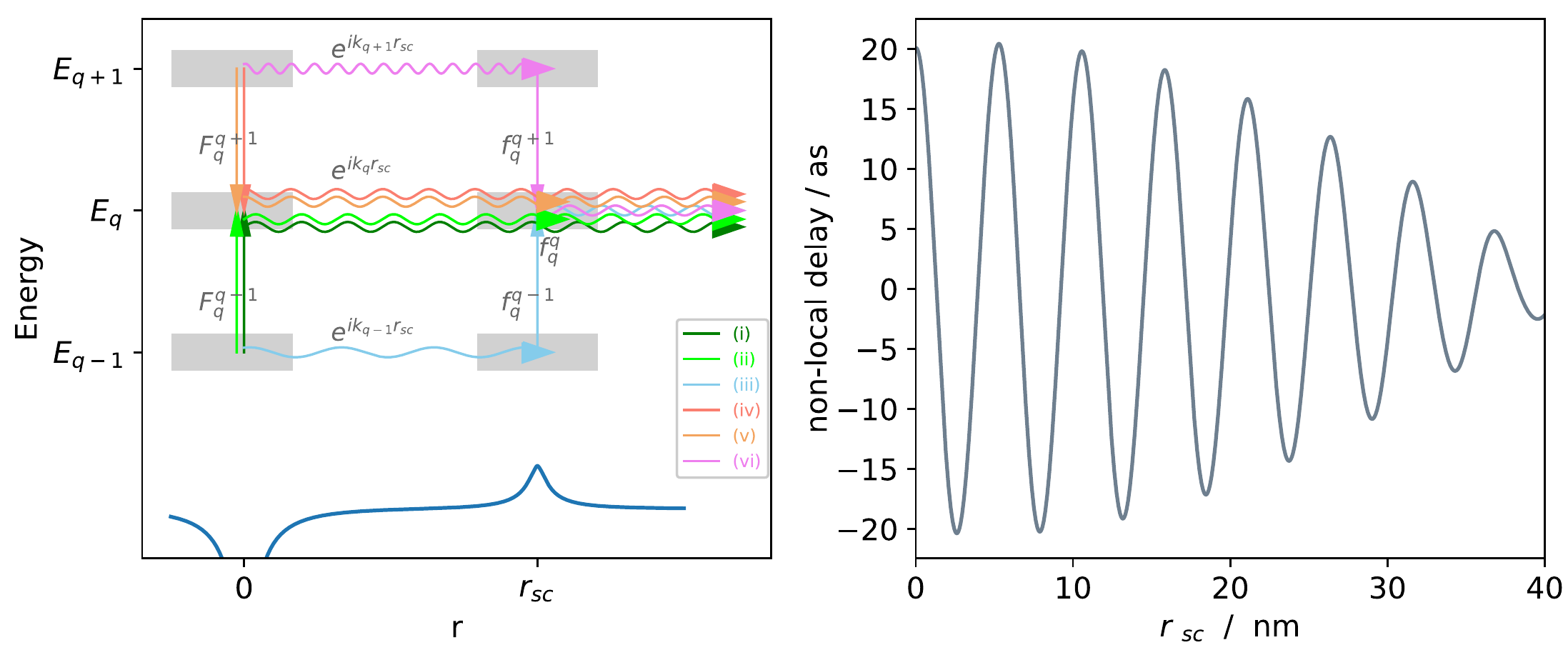}
  \caption{
    Oscillation of the non-local delay $\tau_{\rm nl}$ with the distance $r_{\rm sc}$ of ionization potential to the perturbing potential (right).
    The potential is shown schematically at the bottom of the left panel. Also shown in the left panel are trajectories that can be used to understand the behavior of $\tau_{\rm nl}$ within the soft-photon approximation.
  }
  \label{fig:ForwardsPaths}
\end{figure}

In particular, for the 1D model with one perturbing potential \eqref{eq:yukawa_1D} placed at a distance $r_{\rm sc}$
it was found that the non-local delay oscillates around the local delay as a function of $r_{\rm sc}$ \cite{PhysRevA.97.063415}.
Figure \ref{fig:ForwardsPaths} shows the arrangement of the potentials (left panel) and the non-local delay (right panel).
The frequency of this oscillation as well as the oscillation of the envelope can be explained with a few trajectories within the soft-photon approximation.
Specifically, the sideband intensity obtained by solving the TDSE can be reproduced with the amplitude
\begin{align}
 f_q = F_{q}^{q-1} e^{i k_q r_{\rm sc}} + F_{q}^{q-1} e^{i k_q r_{\rm sc}} f_q^q
     + F_{q-1}^{q-1} e^{i k_{q-1} r_{\rm sc}} f_q^{q-1} 
     + F_{q}^{q+1} e^{i k_q r_{\rm sc}} + F_{q}^{q+1} e^{i k_q r_{\rm sc}} f_q^q
     + F_{q+1}^{q+1} e^{i k_{q+1} r_{\rm sc}} f_q^{q+1} 
   \label{eq:f01}
\end{align}
where the six terms correspond to the paths
\begin{enumerate}[(i)]
\item ionization to state $q-1$, local absorption of a photon to state $q$, and propagation to the scattering site without scattering, 
\item ionization to state $q-1$, local absorption of a photon to state $q$, propagation to the scattering site, and scattering without photon exchange,
\item ionization to state $q-1$, propagation to the scattering site, and scattering with photon absorption to state $q$,
\item ionization to state $q+1$, local emission of a photon to state $q$, and propagation to the scattering site without scattering, 
\item ionization to state $q+1$, local emission of a photon to state $q$, propagation to the scattering site, and scattering without photon exchange, and
\item ionization to state $q+1$, propagation to the scattering site, and scattering with photon emission to state $q$.
\end{enumerate}
These paths are sketched and labeled in the left panel of Figure \ref{fig:ForwardsPaths}.
Paths (i) and (ii), as well as paths (iv) and (v) can be combined into one effective path with amplitude $(1 + f_q^q)$ for scattering at the perturber,
resulting in an overall phase introduced by additional scattering in (ii) and (iv). 
For simplicity, we may also neglect the paths (ii) and (v), which contribute only little, and write the scattering amplitude as
\begin{align}
  \begin{split}
    |f_q|^2
    &= | F_{q}^{q-1} e^{i k_q r_{\rm sc}} + F_{q-1}^{q-1} e^{i k_{q-1} r_{\rm sc}} f_q^{q-1}  + F_{q}^{q+1} e^{i k_q r_{\rm sc}}  + F_{q+1}^{q+1} e^{i k_{q+1} r_{\rm sc}} f_q^{q+1} |^2 \\
    &=  |A|^2 + |B|^2 + 2|A|\,|B| \, \cos\left(-2\delta + \arg(AB^*)\right),
  \end{split}
  \label{eq:f011}
\end{align}
with $A$ and $B$ defined as
\begin{align}
  \begin{split}
    A&=F_{q}^{q-1} e^{i k_q r_{\rm sc}} + F_{q-1}^{q-1} e^{i k_{q-1} r_{\rm sc}} f_q^{q-1} \\
    B&=F_{q}^{q+1} e^{i k_q r_{\rm sc}} + F_{q+1}^{q+1} e^{i k_{q+1} r_{\rm sc}} f_q^{q+1}.
  \end{split}
  \label{eq:f02}
\end{align}
From \eqref{eq:f011} we can see that the phase $\arg(AB^*)$ which provides the scattering delay $\tau$ is
\begin{multline}
    \arg(AB^*)
    = \arg\Big( 
      F_{q  }^{q-1}            {F_{q  }^{q+1}}^*   +  
      F_{q  }^{q-1}            {F_{q+1}^{q+1}}^* {f_q^{q+1}}^* e^{i (k_{q  }-k_{q+1}) r_{\rm sc}} +  \\ 
      F_{q-1}^{q-1}  f_q^{q-1} {F_{q  }^{q+1}}^*               e^{i (k_{q-1}-k_{q  }) r_{\rm sc}} + 
      F_{q-1}^{q-1}  f_q^{q-1} {F_{q+1}^{q+1}}^* {f_q^{q+1}}^* e^{i (k_{q-1}-k_{q+1}) r_{\rm sc}}  \Big).
  \label{eq:f03}
\end{multline}
The first term of \eqref{eq:f03}, independent of the perturber distance, defines the local (photoionization) phase.
This term, typically the largest in magnitude, is modulated by other terms, which rotate in the complex plane with increasing $r_{\rm sc}$, with rotation frequencies
$k_{q+1}-k_{q-1}$, $k_{q+1}-k_{q}$ and $k_{q}-k_{q-1}$. We will refer to these oscillations as ``slow'', as we will discuss  more rapid oscillations below.
Similar conclusions may be derived from perturbation theory\cite{dahlstrom2012}.

We note, however, that both the LAPE amplitude $F^{q+\nu}_{q}$ and the LAES amplitude $f^{q+\nu}_{q}$ have a direction-dependence which reduces to a forward-backward dependence in one dimension.
In Eq. \eqref{eq:f011} and in Ref. \cite{PhysRevA.97.063415} only forward scattering is discussed, while we additionally consider backward scattering below.
For LAES the magnitude of the amplitudes can differ for forward and backward scattering due to the momentum difference in the argument of the Bessel function in \eqref{eq:laes}.
With the parameters of our model (which correspond to experimental parameters), photon absorption or emission during scattering is considerably more likely to happen when the electron is scattered backwards than when it is scattered in forward direction, hence trajectories that are back-scattered can contribute in some arrangements, as discussed below.

\section{\label{sec:results}Results}

\begin{figure}[!h]
  \includegraphics[width=0.9\textwidth]{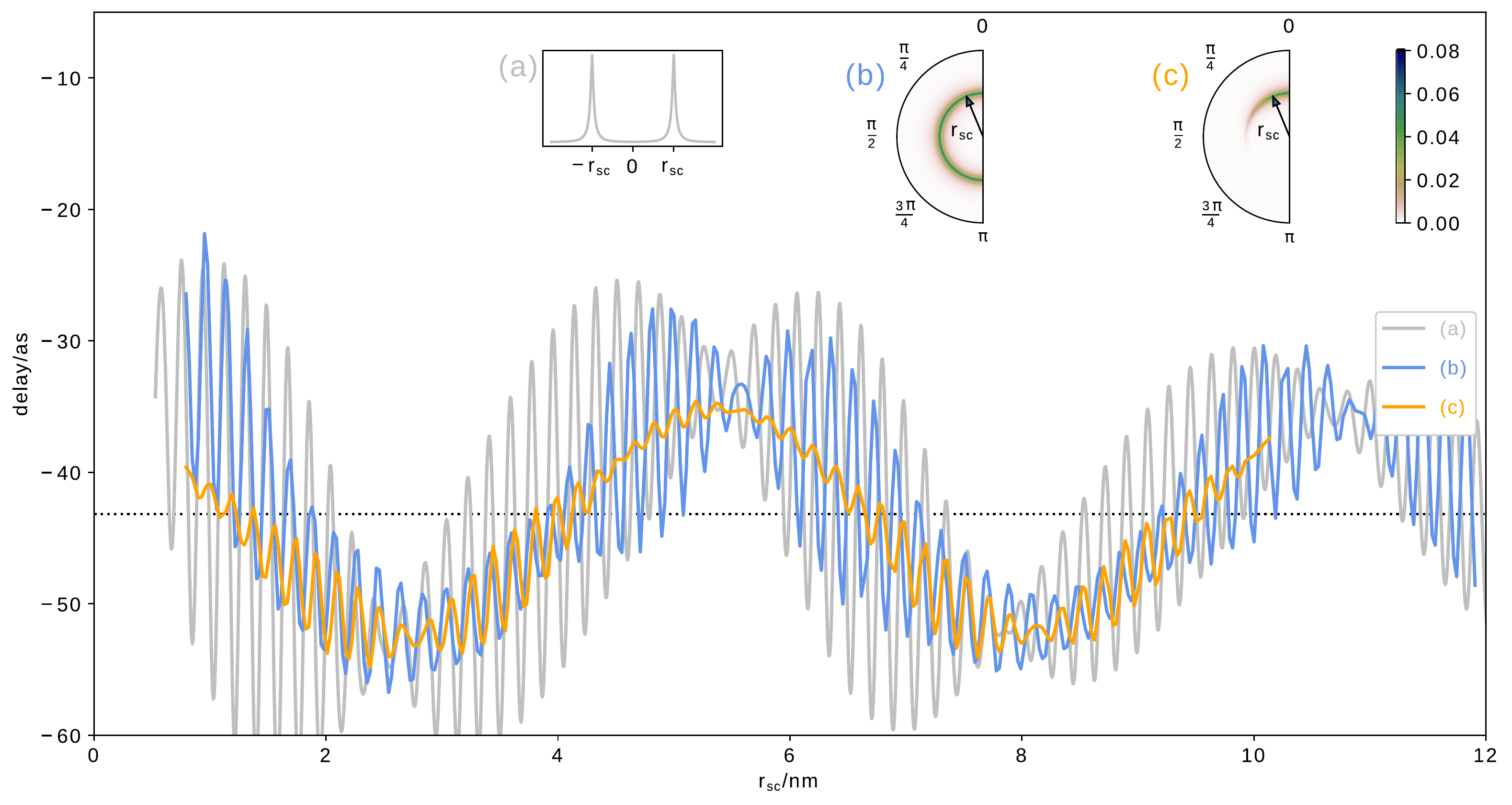}
  \caption{
    Total time delay as function of the  distance  from the point of ionization to the center of the perturbing potential. Inset (a) shows the symmetric 1D perturbing potentials,
    insets (b) and (c) show the cross cuts of the 3D perturbing potentials through any plane containing the $\hat{\bm{z}}$ axis, around which they are symmetrical. The detector is located in $\theta=0$ direction. 
    The graph shows calculated delays for each potential:
    A gray line for the 1D potential (a), a blue line for fully spherically symmetrical potential (b)
    and an orange line for the potential decreasing from  $\theta=0$ to $\theta=\pi$ (c).
  }
  \label{fig:rabbitSphPot}
\end{figure}

\begin{figure}[!h]
  \includegraphics[width=0.9\textwidth]{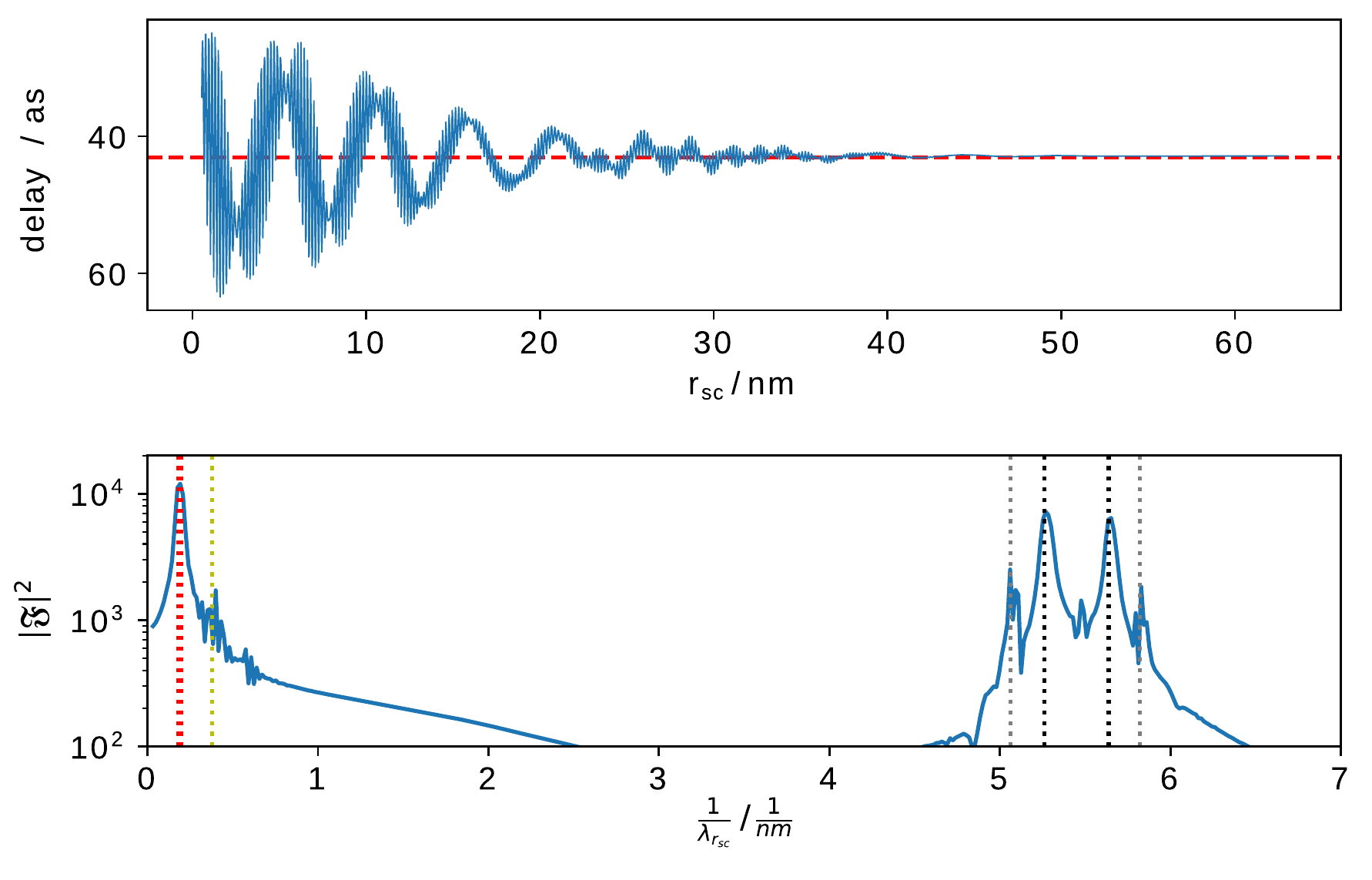}
  \caption{
    Time delay as function of $r_{\rm sc}$, for the 1D TDSE calculation, with two perturbing potentials located symmetrically with respect to the binding potential at the origin.
    Top: Time delay.
    Bottom: Fourier transform of the time delay as a function of $r_{\rm sc}$, where  $\lambda_{\rm r_{ sc}}$ is the spatial frequency. Gray and black horizontal dotted lines between \unit[5]{nm$^{-1}$} and \unit[6]{nm$^{-1}$} correspond to the spatial frequencies in  Figure \ref{fig:1dPaths} below. 
    The two  red vertical lines are positioned at
    $\lambda_{\rm B}(k_{q+1}-k_{q})$ and $\lambda_{\rm B}(k_{q}-k_{q-1})$, which are very close and not resolved on the current scale.
    The yellow vertical dotted line is located at $\lambda_{\rm B}(k_{q+1}-k_{q-1})$.
  }
  \label{fig:Fourier1D_TDSE}
\end{figure}

In the following, the non-local delay in three spatial dimensions is investigated. 
To understand some of the results, however, we first turn again to the 1D model described in section \ref{subsec:1D_calculations} and used in \cite{PhysRevA.97.063415}, but with a symmetric arrangement of the perturbing potentials:
The centers of two scattering potentials $V_{\rm p}$ are located symmetrically, at the same distance $r_{\rm sc}$, from the ionization site, as shown in inset (a) of Figure \ref{fig:rabbitSphPot}.
The corresponding total delay, obtained from solving the TDSE, is shown as gray line in the figure.
An oscillation of the delay with $r_{\rm sc}$ similar to scattering at one potential $V_{\rm p}$ is observed, with a period of \unit[5.3]{nm}.
The same calculation shows an effective decay on a larger scale of $r_{\rm sc}$, as can be seen in  upper panel Figure \ref{fig:Fourier1D_TDSE}, which is a consequence of the finite distance that an electron can reach during pulse propagation. 

These are oscillations of $\tau_{\rm nl}$ around the constant value of the local delay \unit[$\tau_{\rm l} = -43$]{as}.
Additionally, there are also oscillations visible with a much shorter period of ca.\ \unit[0.18]{nm}.
The difference of the amplitudes of ``fast'' oscillations suggests  that multiple oscillations frequencies are involved, which is confirmed by taking the Fourier transform of the total delay.
As shown in the bottom panel of Figure \ref{fig:Fourier1D_TDSE}, the fast oscillation is composed of two oscillations that are centered in between the de Broglie half-wavelengths of the electrons in states $15$, $16$, and $17$.
The same figure also shows the frequencies of the  ``slow'' oscillations.

\begin{figure}[!h]
  \includegraphics[width=0.9\textwidth]{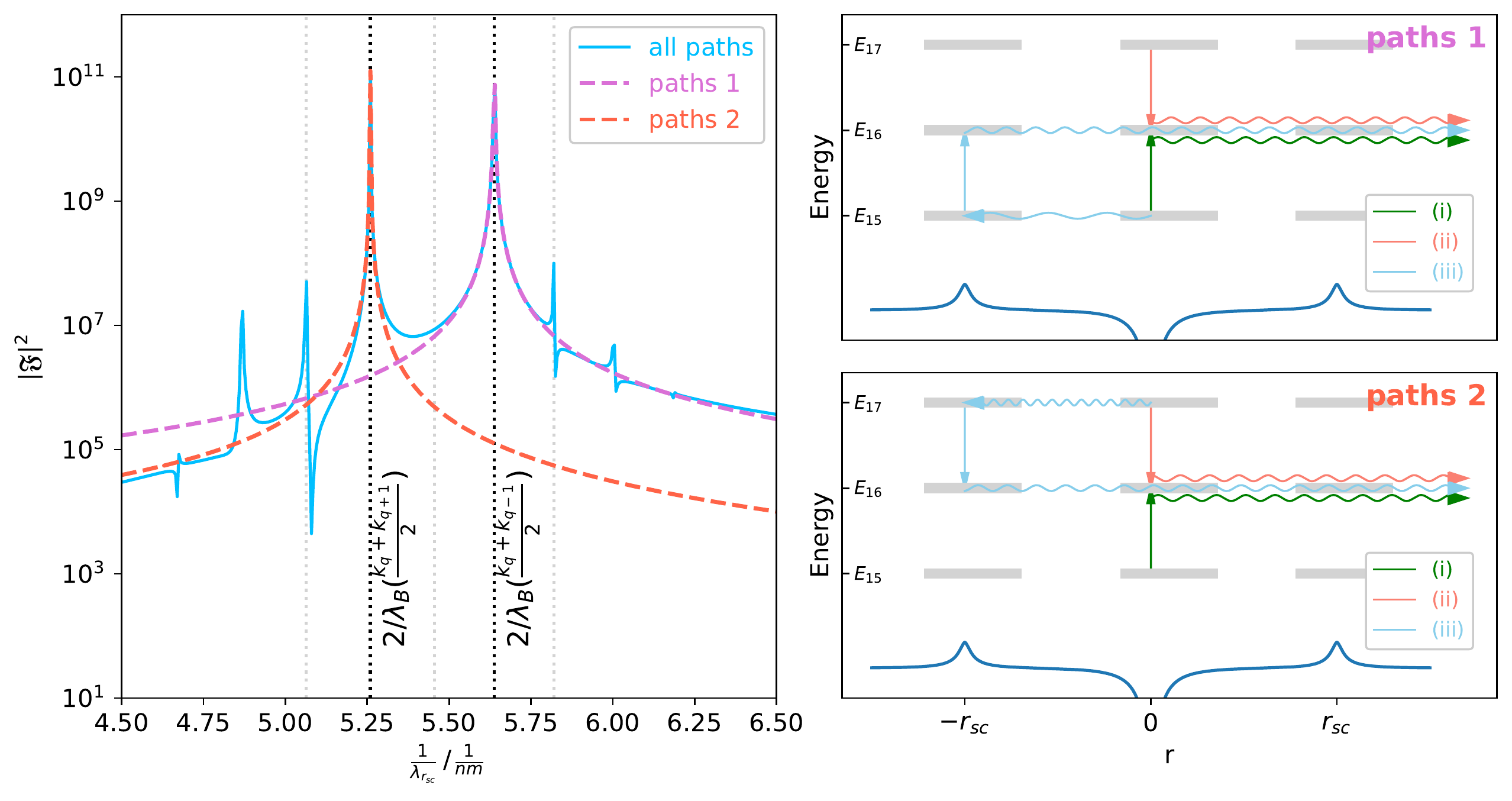}
  \caption{
    Left: Fourier transform of the time delay as function of $r_{\rm sc}$, for the 1D model with two perturbing potentials located symmetrically with respect to the binding potential at origin.
    The time delay is calculated using the trajectory model described in the text, including all trajectories (blue full line), including only trajectories schematically indicated in the right-hand upper panel (dashed violet line with peak at ca.\ \unit[5.65]{nm$^{-1}$}), and including only trajectories shown in the right-hand lower panel (dashed red line with peak at ca.\ \unit[5.25]{nm$^{-1}$}).
    The vertical lines indicate the de Broglie half-wavelengths of electrons in the states 15, 16, and 17.
    Right: Sketch of the photoelectron paths used in the left-hand side panel.
  }
  \label{fig:1dPaths}
\end{figure}

\begin{figure}[!h]
  \includegraphics[width=0.9\textwidth]{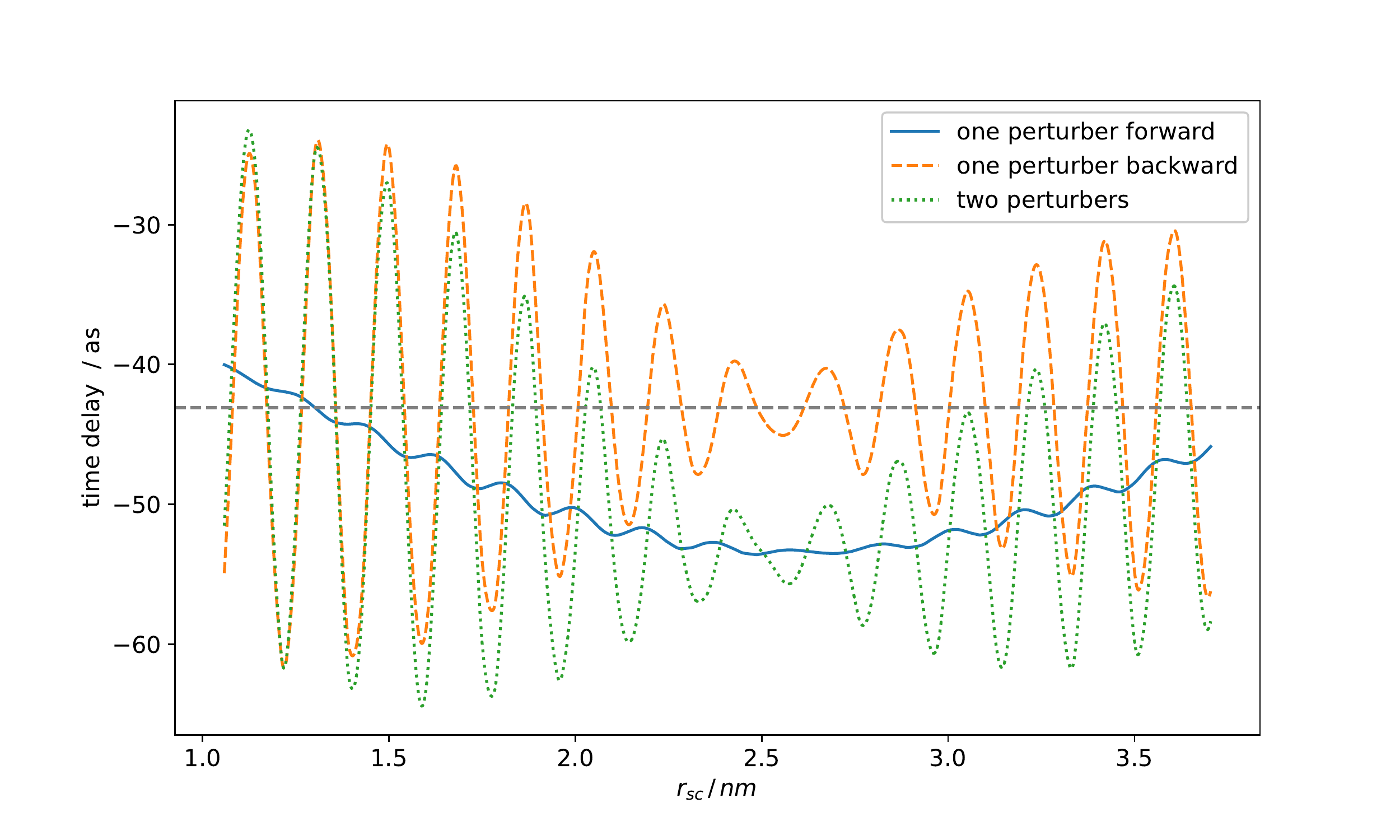}
  \caption{
    Total time-delay for solutions of the one-dimensional model with one scattering potential at $r_{\rm sc}$ and detection of forward-scattered trajectories (``one perturber, forward''), with one scattering potential at $r_{\rm sc}$ and detection of back-scattered trajectories  (``one perturber, backwards''), and with two scattering potentials at $x = \pm r_{\rm sc}$ (``two perturbers'').
    }
  \label{fig:1dForwBack}
\end{figure}

The origin of the oscillatory behavior of the non-local time delay can be well explained with the trajectory model from section \ref{subsec:nonlocal_paths}.
The relevant paths which contribute to ``slow'' oscillations are illustrated in the right panel of Figure \ref{fig:ForwardsPaths}.
While the previously reported results explain the slow oscillations, we extend the trajectory model in order to explain ``fast'' oscillations.
Therefore, in the arrangement with two symmetrical potentials, we allow electron trajectories to propagate in the negative direction and to scatter back at the left perturber towards the detection direction.
We restrict our model to all possible paths that end in state $q$, that have exchanged one photon with the assisting laser field, and that scattered at most once.
The paths are illustrated on the right-hand side of Figure \ref{fig:1dPaths}, where the top panel shows three paths:
The two local pathways representing
\begin{enumerate}[(i)]
\item   photoionization to state $q-1$, local absorption to state $q$, and motion to the detector in positive $r$-direction, as well as
\item   photoionization to state $q+1$, local emission to state $q$, and motion to the detector in positive $r$-direction, 
\end{enumerate}
and the non-local pathway 
\begin{enumerate}[(i)] \setcounter{enumi}{2}
\item   photoionization to state $q-1$, motion in negative $r$-direction, backscattering at the perturber with photon absorption to state $q$, and motion to the detector in positive $r$-direction.
\end{enumerate}
The bottom panel shows similar pathways but with the non-local pathway originating from photon emission from state $q+1$, rather than absorption from state $q-1$.
The selected pathways are contributing to the transition amplitude in a coherent sum that can be written in the same way  as \eqref{eq:f011}, i.e.

\begin{equation}
  f= 
  F^{q-1}_{q^{(+)}} e^{i\left(-\delta+k_q r_{sc} \right)} + F^{q+1}_{q^{(+)}} e^{i\left(\delta+k_q r_{sc} \right)}   +
  \begin{cases}
    F^{q-1}_{q^{(-)}} e^{i\left(-\delta+ \left(k_{q-1}+2k_q \right)  r_{sc} \right)}f^{q-1^{(-)}}_{q^{(+)}},  & \text{paths 1 }  \\
    F^{q+1}_{q^{(-)}} e^{i\left(\delta+ \left(k_{q+1}+2k_q \right)  r_{sc} \right)}f^{q+1^{(-)}}_{q^{(+)}},  & \text{paths 2 }.   %
  \end{cases}
  \label{eq:paths1}
\end{equation}
For the amplitudes, we added the direction in which the electron moves after ionization or before and after scattering as superscript and subscript $(\pm)$, respectively, indicating motion into positive or negative $x$-direction.
From \eqref{eq:paths1}, the two sets of trajectories, paths 1 and paths 2, taken independently, result in an oscillation of the non-local delay with frequencies $(k_{q}+k_{q+1})/2$  and $(k_{q}+k_{q-1})/2$, respectively. 
Taking a coherent superposition of the two sets of trajectories, we see the appearance of additional oscillation frequencies, but much smaller in amplitude compared to the main peaks. 
In general, the magnitude of the oscillatory behavior is determined by the absolute value of the products of transition amplitudes for each oscillatory factor in \eqref{eq:f03}, while the phase will be directly reflected in the measured time delay.
We can see that already the four paths used in this model can correctly predict the oscillatory behavior introduced by the additional scattering potential on the left-hand side from the binding potential, i.e., the peaks marked by black and gray dotted horizontal lines in the left panel if Figure \ref{fig:1dPaths} correspond exactly to the position of the  peaks in the TDSE calculation, shown in the bottom panel of Figure \ref{fig:Fourier1D_TDSE}. 
The fast oscillations are a result of backscattering, therefore they can be seen even in a setup with one scattering potential but with the photoelectron flux detected on the opposite side from the scattering potential with respect to the binding potential. In Figure \ref{fig:1dForwBack} we compare a 1D TDSE calculation of the non-local delay for three different cases: One perturber and detection of forward-scattered electrons (blue full line), one perturber and backward scattered electrons detected (orange dashed line), and two perturbers (green dotted line).
The backscattering is responsible for the ``fast'' oscillations (orange and green line) while the ``slow'' oscillations are absent in the case without forward scattering, as predicted by the trajectory model.
We also note that there is an additional ``fast'' oscillation with small amplitude for the case of forward-scattering at one perturber coming from the shape of the perturbing potential. 

Next, we study AI with scattering for the 3D hydrogen atom.
Our first test case is the 3D equivalent of the 1D case, i.e., from the 1D scattering potential \eqref{eq:yukawa_1D} we construct the corresponding shell-like spherically symmetric potential
\begin{equation}
  V_{\rm p}(r,r_{\rm sc}) = V_0 \frac{ e^{-\frac{\left|r-r_{\rm sc}\right|}{\lambda}}  }{\sqrt{\left(r-r_{\rm sc}\right)^2+s^2}},
  \label{eq:yukawa_semi1D}
\end{equation}
which is shown as polar plot in inset (b) in Figure \ref{fig:rabbitSphPot}. 
The resulting time delay measured in $\theta = 0$ direction (i.e., in the polarization direction of the laser), plotted as blue line in the figure, shows similar oscillations with $r_{\rm sc}$ like as 1D case. 
We find the relatively slow oscillation that is also present for the 1D model with only one perturbing potential, as well as the fast oscillations that are due to backscattering.
The oscillations have the same frequency composition  as in the 1D case. However, the fast oscillations are weaker in amplitude, suggesting a different scattering cross section.
We attribute the weaker amplitude to the reduced symmetry of the process, as the potentials are spherically symmetric but the overall problem is only cylindrically symmetric due to the laser field.
A different way of understanding the smaller amplitude
is that there is a larger ``phase space'' available for the trajectories and a smaller fraction of them interferes,
while a large part changes to higher angular momenta without contributing to the interference.
Interpreted in a path picture, there is interference of forward-scattered paths with other paths scattered at different angles and experiencing a different potential and thus a different scattering phase shift, leading to a reduction of the oscillation contrast relative to the 1D case.

Next, we modify the 3D scattering potential to reduce the back-scattering contribution, i.e., we introduce an angular dependence by multiplying the potential \eqref{eq:yukawa_semi1D} with the
angle-dependent function $V=V_{\rm p}(r,r_{\rm sc})\alpha(\theta)$ where $\alpha(\theta)=\frac{\cos(\theta)+1}{2} e^{-\theta^2}$.
The resulting potential is shown as inset (c) in Figure \ref{fig:rabbitSphPot} and the corresponding delay is shown as orange line in the figure.
Due to its shape, we call this potential the half-shell potential.
The delay for the half-shell potential has visible oscillations with the distance $r_{\rm sc}$ around the value of the local delay, too, but the amplitudes are reduced compared to those of the shell potential.
The less-pronounced fast oscillation is due to a much smaller back-scattering contribution, which, however, still exist, as the perturbing potential is exactly zero only for $\theta=\pi$.
The less-pronounced slow oscillation is due to  
the fact that the magnitude of the scattering potential is now angle-dependent.

\begin{figure}[!h]
  \includegraphics[width=0.9\textwidth]{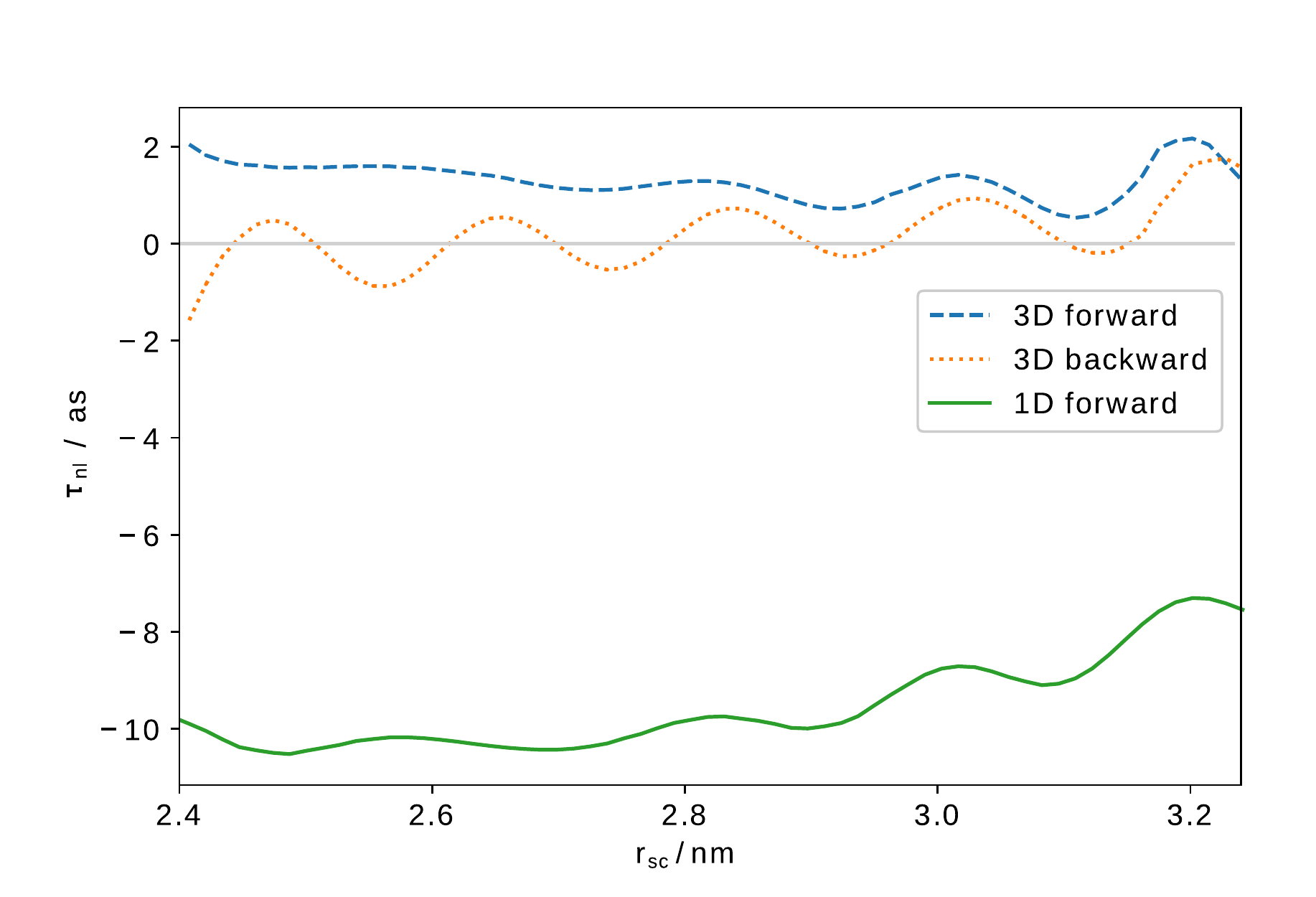}
  \caption{
    Comparison of the non-local delay obtained in 1D and 3D simulations. ``1D forward'' and ``3D forward'' represent scattering at one scattering center and measurement of forward scattering, while ``3D backward'' represents scattering at one scattering center and measurement of the back-scattered contribution.
  }
  \label{fig:delay1dVs3D}
\end{figure}

The considered shell and half-shell potentials are not the typical scattering potentials for electron transport in condensed matter.
Thus, we consider a spherically symmetrical Yukawa potential shifted from the origin to the positive side of the $\hat{\bm{z}}$-axis, defined by
\begin{equation}
  V_{\rm p}(\bm{r},r_{\rm sc}) = V_0
  \frac{
    \exp^{  \left(-\frac{\left|\bm{r}-r_{\rm sc}\hat{\bm{z}}\right|}{\lambda}\right)  }
  }
  {
    \sqrt{\left(\bm{r}-r_{\rm sc}\hat{\bm{z}}\right)^2+s^2}.
  },
  \label{eq:yukawa_3D}
\end{equation}
In contrast to the previous cases, for this potential we find only very weak oscillations of the delay with $r_{\rm sc}$.
Viewed again in a trajectory model, the weaker oscillations are a consequence of averaging over the delays of all possible electron trajectories that reach the detector, or, equivalently, of a reduced forward-scattering amplitude.
In contrast to the full-shell potential, where all trajectories always propagate through a potential barrier of the same height, the potential height for different trajectories meeting such a potential varies from 0 to $V_0$, implying a different scattering phase shift for each of them.
This effect appears to be sufficient to reduce the non-local delay contributions to a value of a few attoseconds at most (Figure \ref{fig:delay1dVs3D}).
In Figure \ref{fig:delay1dVs3D}, the 3D time delay is compared with the 1D results. The results are shown relative to the local delay. 
We notice that the slow oscillations are reduced by almost an order of magnitude for forward scattering detection.
The back-scattering shows clear fast oscillations compared to forward scattering. This is due to the additional scattering on the binding potential and consequently the formation of a standing wave.
Compared to the 1D case with two symmetrical perturbers (gray line in Fig. \ref{fig:rabbitSphPot}), we can see that magnitude of these fast  oscillations is significantly reduced.

\section{\label{sec:conclusion}Conclusion}

In this article, we investigated the influence of scattering centers on the photoionization time delay as a model system for AI experiments in condensed matter.
A previous study with one-dimensional model potentials showed that scattering can change the time delay and that the total delay is a sum of the local delay,
which encodes the electronic structure of the target, and a non-local delay, which encodes the scattering properties.
The distance to the scattering center is important in the 1D case as the magnitude of the delay is sensitive to it.
All effects can be explained with the help of a simple trajectory model.

Here, we have extended this 1D study by considering a symmetric arrangement of the scattering potentials, which reveals additional dependencies of the delay on the distance to the scattering center due to backscattering.
Similar results have been obtained for a 3D hydrogen atom with scattering of the ionized electron at a shell-like potential with spherical symmetry.
However, if the symmetry is reduced by turning the spherical shell potential into a half-shell potential, we find less and less modulation contrast in the non-local delay.
For the case of scattering at a perturber centered around a point remote from the origin the contribution of the non-local delay is signicantly reduced.

The previous study \cite{PhysRevA.97.063415} has also shown that collisions with multiple randomly distributed perturbing potentials lead to a decay of the non-local delay with the mean free path,
thus making the non-local delay negligible if the mean free path is long.
Taken together with our results presented here, we conclude that experiments in disordered systems like liquid phases
at the low photon energies used in this work, are unlikely to be sensitive to scattering during electron transport.
Thus, AI for such systems is expected to mainly be sensitive to changes of the electronic structure of the ionized molecule due to its neighbors, but not to scattering of the photoelectron at other molecules.
This is a very important insight, as it helps to analyze AI experiments correctly.
This agrees with the interpretation of our experimental results presented in \cite{IJordanSci2020}.

However, for systems with long-range order the non-local effect might be interesting to study further, as the symmetry can lead to interference effects that may provide structural information from AI measurements.
Also our seemingly artificial shell and half-shell potentials can serve as a model for atoms and molecules in cavities, such as endohedral fullerenes or other molecular cages,
as discussed in ~\cite{PhysRevA.80.011201}.
As the delay is sensitive to the size of the shell, our calculations show that AI may be a suitable tool to study such systems.


\section{\label{sec:Acknowledgements}Acknowledgements}

This project received funding from the European Union’s Horizon 2020 research and innovation programme under the Marie Sk\l{}odowska-Curie grant agreement No 801459 - FP-RESOMUS - and the Swiss National Science Foundation through the NCCR MUST. AS is grateful for support by an Ambizione grant of the Swiss National Science Foundation.

\bibliography{literature}

\begin{thebibliography}{10}
\providecommand{\url}[1]{#1}
\csname url@samestyle\endcsname
\providecommand{\newblock}{\relax}
\providecommand{\bibinfo}[2]{#2}
\providecommand{\BIBentrySTDinterwordspacing}{\spaceskip=0pt\relax}
\providecommand{\BIBentryALTinterwordstretchfactor}{4}
\providecommand{\BIBentryALTinterwordspacing}{\spaceskip=\fontdimen2\font plus
\BIBentryALTinterwordstretchfactor\fontdimen3\font minus
  \fontdimen4\font\relax}
\providecommand{\BIBforeignlanguage}[2]{{%
\expandafter\ifx\csname l@#1\endcsname\relax
\typeout{** WARNING: IEEEtran.bst: No hyphenation pattern has been}%
\typeout{** loaded for the language `#1'. Using the pattern for}%
\typeout{** the default language instead.}%
\else
\language=\csname l@#1\endcsname
\fi
#2}}
\providecommand{\BIBdecl}{\relax}
\BIBdecl

\bibitem{klunder2011}
\BIBentryALTinterwordspacing
K.~Kl\"under, J.~M. Dahlstr\"om, M.~Gisselbrecht, T.~Fordell, M.~Swoboda,
  D.~Gu\'enot, P.~Johnsson, J.~Caillat, J.~Mauritsson, A.~Maquet, Ta\"{i}eb,
  and A.~L'Huillier, ``Probing single-photon ionization on the attosecond time
  scale,'' \emph{Phys. Rev. Lett.}, vol. 106, p. 143002, Apr 2011. [Online].
  Available: \url{http://link.aps.org/doi/10.1103/PhysRevLett.106.143002}
\BIBentrySTDinterwordspacing

\bibitem{guenot2012}
\BIBentryALTinterwordspacing
D.~Gu\'enot, K.~Kl\"under, C.~L. Arnold, D.~Kroon, J.~M. Dahlstr\"om,
  M.~Miranda, T.~Fordell, M.~Gisselbrecht, P.~Johnsson, J.~Mauritsson,
  E.~Lindroth, A.~Maquet, R.~Ta\"{\i}eb, A.~L'Huillier, and A.~S. Kheifets,
  ``{Photoemission-time-delay measurements and calculations close to the
  3$s$-ionization-cross-section minimum in Ar},'' \emph{Phys. Rev. A}, vol.~85,
  p. 053424, May 2012. [Online]. Available:
  \url{http://link.aps.org/doi/10.1103/PhysRevA.85.053424}
\BIBentrySTDinterwordspacing

\bibitem{guenot2014}
\BIBentryALTinterwordspacing
D.~Gu{\'e}not, D.~Kroon, E.~Balogh, E.~W. Larsen, M.~Kotur, M.~Miranda,
  T.~Fordell, P.~Johnsson, J.~Mauritsson, M.~Gisselbrecht, K.~Varj{\`u}, C.~L.
  Arnold, T.~Carette, A.~S. Kheifets, E.~Lindroth, A.~L'Huillier, and J.~M.
  Dahlstr{\"o}m, ``Measurements of relative photoemission time delays in noble
  gas atoms,'' \emph{Journal of Physics B: Atomic, Molecular and Optical
  Physics}, vol.~47, no.~24, p. 245602, 2014. [Online]. Available:
  \url{http://stacks.iop.org/0953-4075/47/i=24/a=245602}
\BIBentrySTDinterwordspacing

\bibitem{palatchi2014}
\BIBentryALTinterwordspacing
C.~Palatchi, J.~M. Dahlstr\"om, A.~S. Kheifets, I.~A. Ivanov, D.~M. Canaday,
  P.~Agostini, and L.~F. DiMauro, ``Atomic delay in helium, neon, argon and
  krypton,'' \emph{Journal of Physics B: Atomic, Molecular and Optical
  Physics}, vol.~47, no.~24, p. 245003, 2014. [Online]. Available:
  \url{http://stacks.iop.org/0953-4075/47/i=24/a=245003}
\BIBentrySTDinterwordspacing

\bibitem{heuser2016}
\BIBentryALTinterwordspacing
S.~Heuser, A.~Jim\'enez~Gal\'an, C.~Cirelli, C.~Marante, M.~Sabbar, R.~Boge,
  M.~Lucchini, L.~Gallmann, I.~Ivanov, A.~S. Kheifets, J.~M. Dahlstr\"om,
  E.~Lindroth, L.~Argenti, F.~Mart\'{\i}n, and U.~Keller, ``Angular dependence
  of photoemission time delay in helium,'' \emph{Phys. Rev. A}, vol.~94, p.
  063409, Dec 2016. [Online]. Available:
  \url{https://link.aps.org/doi/10.1103/PhysRevA.94.063409}
\BIBentrySTDinterwordspacing

\bibitem{huppert2016}
M.~Huppert, I.~Jordan, D.~Baykusheva, A.~{von Conta}, and H.~J. W\"orner,
  ``Attosecond delays in molecular photoionization,'' \emph{Phys. Rev. Lett.},
  vol. 117, no. 093001, 2016.

\bibitem{jordan2017}
I.~Jordan, M.~Huppert, S.~Pabst, A.~S. Kheifets, D.~Baykusheva, and H.~J.
  W{\"o}rner, ``Spin-orbit delays in photoemission,'' \emph{Phys Rev A},
  vol.~95, no. 013404, 2017.

\bibitem{tao2016}
Z.~Tao, C.~Chen, T.~Szilv\'asi, M.~Keller, M.~Mavrikakis, H.~Kapteyn, and
  M.~Murnane, ``Direct time-domain observation of attosecond final-state
  lifetimes in photoemission from solids,'' \emph{Science}, vol. 353, no. 6294,
  pp. 62--67, 2016.

\bibitem{siek2017angular}
F.~Siek, S.~Neb, P.~Bartz, M.~Hensen, C.~Str{\"u}ber, S.~Fiechter,
  M.~Torrent-Sucarrat, V.~M. Silkin, E.~E. Krasovskii, N.~M. Kabachnik
  \emph{et~al.}, ``Angular momentum--induced delays in solid-state
  photoemission enhanced by intra-atomic interactions,'' \emph{Science}, vol.
  357, no. 6357, pp. 1274--1277, 2017.

\bibitem{jordan2015photoelectron}
I.~Jordan, M.~Huppert, M.~Brown, J.~A. van Bokhoven, and H.~J. W{\"o}rner,
  ``Photoelectron spectrometer for attosecond spectroscopy of liquids and
  gases,'' \emph{Review of Scientific Instruments}, vol.~86, no.~12, p. 123905,
  2015.

\bibitem{IJordanSci2020}
\BIBentryALTinterwordspacing
I.~Jordan, M.~Huppert, D.~Rattenbacher, M.~Peper, D.~Jelovina, C.~Perry, A.~von
  Conta, A.~Schild, and H.~J. W{\"o}rner, ``Attosecond spectroscopy of liquid
  water,'' \emph{Science}, vol. 369, no. 6506, pp. 974--979, 2020. [Online].
  Available: \url{https://science.sciencemag.org/content/369/6506/974}
\BIBentrySTDinterwordspacing

\bibitem{kheifets2010}
\BIBentryALTinterwordspacing
A.~S. Kheifets and I.~A. Ivanov, ``Delay in atomic photoionization,''
  \emph{Phys. Rev. Lett.}, vol. 105, no.~23, p. 233002, 2010. [Online].
  Available: \url{http://link.aps.org/doi/10.1103/PhysRevLett.105.233002}
\BIBentrySTDinterwordspacing

\bibitem{nagele2011}
\BIBentryALTinterwordspacing
S.~Nagele, R.~Pazourek, J.~Feist, K.~Doblhoff-Dier, C.~Lemell, K.~Tok\'esi, and
  J.~Burgd\"orfer, ``Time-resolved photoemission by attosecond streaking:
  extraction of time information,'' \emph{Journal of Physics B: Atomic,
  Molecular and Optical Physics}, vol.~44, no.~8, pp. 081\,001--, 2011.
  [Online]. Available: \url{http://stacks.iop.org/0953-4075/44/i=8/a=081001}
\BIBentrySTDinterwordspacing

\bibitem{pazourek2012}
\BIBentryALTinterwordspacing
R.~Pazourek, J.~Feist, S.~Nagele, and J.~Burgd\"orfer, ``Attosecond streaking
  of correlated two-electron transitions in helium,'' \emph{Phys. Rev. Lett.},
  vol. 108, p. 163001, Apr 2012. [Online]. Available:
  \url{http://link.aps.org/doi/10.1103/PhysRevLett.108.163001}
\BIBentrySTDinterwordspacing

\bibitem{dahlstrom2012}
\BIBentryALTinterwordspacing
J.~M. Dahlstr\"om, A.~L'Huillier, and A.~Maquet, ``Introduction to attosecond
  delays in photoionization,'' \emph{Journal of Physics B: Atomic, Molecular
  and Optical Physics}, vol.~45, no.~18, p. 183001, 2012. [Online]. Available:
  \url{http://stacks.iop.org/0953-4075/45/i=18/a=183001}
\BIBentrySTDinterwordspacing

\bibitem{ivanov2012}
\BIBentryALTinterwordspacing
I.~A. Ivanov, A.~S. Kheifets, and V.~V. Serov, ``Attosecond time-delay
  spectroscopy of the hydrogen molecule,'' \emph{Phys. Rev. A}, vol.~86, p.
  063422, Dec 2012. [Online]. Available:
  \url{http://link.aps.org/doi/10.1103/PhysRevA.86.063422}
\BIBentrySTDinterwordspacing

\bibitem{dahlstrom2013}
\BIBentryALTinterwordspacing
J.~Dahlstr\"om, D.~Gu\'enot, K.~Kl\"under, M.~Gisselbrecht, J.~Mauritsson,
  A.~L'Huillier, A.~Maquet, and R.~Ta\"{i}eb, ``Theory of attosecond delays in
  laser-assisted photoionization,'' \emph{Chemical Physics}, vol. 414, no.~0,
  pp. 53--64, Mar. 2013. [Online]. Available:
  \url{http://www.sciencedirect.com/science/article/pii/S0301010412000298}
\BIBentrySTDinterwordspacing

\bibitem{kheifets2013}
\BIBentryALTinterwordspacing
A.~S. Kheifets, ``Time delay in valence-shell photoionization of noble-gas
  atoms,'' \emph{Phys. Rev. A}, vol.~87, no.~6, pp. 063\,404--, 2013. [Online].
  Available: \url{http://link.aps.org/doi/10.1103/PhysRevA.87.063404}
\BIBentrySTDinterwordspacing

\bibitem{maquet2014}
\BIBentryALTinterwordspacing
A.~Maquet, J.~Caillat, and R.~Ta\"{i}eb, ``Attosecond delays in
  photoionization: time and quantum mechanics,'' \emph{Journal of Physics B:
  Atomic, Molecular and Optical Physics}, vol.~47, no.~20, p. 204004, 2014.
  [Online]. Available: \url{http://stacks.iop.org/0953-4075/47/i=20/a=204004}
\BIBentrySTDinterwordspacing

\bibitem{hockett2016}
\BIBentryALTinterwordspacing
P.~Hockett, E.~Frumker, D.~M. Villeneuve, and P.~B. Corkum, ``Time delay in
  molecular photoionization,'' \emph{Journal of Physics B: Atomic, Molecular
  and Optical Physics}, vol.~49, no.~9, p. 095602, 2016. [Online]. Available:
  \url{http://stacks.iop.org/0953-4075/49/i=9/a=095602}
\BIBentrySTDinterwordspacing

\bibitem{baykusheva2017a}
\BIBentryALTinterwordspacing
D.~Baykusheva and H.~J. W\"{o}rner, ``Theory of attosecond delays in molecular
  photoionization,'' \emph{The Journal of Chemical Physics}, vol. 146, no.~12,
  p. 124306, 2017. [Online]. Available:
  \url{http://dx.doi.org/10.1063/1.4977933}
\BIBentrySTDinterwordspacing

\bibitem{baykusheva2017b}
\BIBentryALTinterwordspacing
D.~Baykusheva and H.~J. W{\"o}rner, ``Comment on `time delays in molecular
  photoionization','' \emph{Journal of Physics B: Atomic, Molecular and Optical
  Physics}, vol.~50, no.~7, p. 078002, 2017. [Online]. Available:
  \url{http://stacks.iop.org/0953-4075/50/i=7/a=078002}
\BIBentrySTDinterwordspacing

\bibitem{itatani2002}
J.~Itatani, F.~Qu{\'e}r{\'e}, G.~L. Yudin, M.~Y. Ivanov, F.~Krausz, and P.~B.
  Corkum, ``Attosecond streak camera,'' \emph{Phys. Rev. Lett.}, vol.~88, p.
  173903, 2002.

\bibitem{kienberger2004}
R.~Kienberger, E.~Goulielmakis, M.~Uiberacker, A.~Baltuska, V.~Yakovlev,
  F.~Bammer, A.~Scrinzi, T.~Westerwalbesloh, U.~Kleineberg, U.~Heinzmann,
  M.~Drescher, and F.~Krausz, ``Atomic transient recorder,'' \emph{Nature},
  vol. 427, pp. 817--821, 2004.

\bibitem{cavalieri2007}
\BIBentryALTinterwordspacing
A.~L. Cavalieri, N.~Muller, T.~Uphues, V.~S. Yakovlev, A.~Baltuska, B.~Horvath,
  B.~Schmidt, L.~Blumel, R.~Holzwarth, S.~Hendel, M.~Drescher, U.~Kleineberg,
  P.~M. Echenique, R.~Kienberger, F.~Krausz, and U.~Heinzmann, ``Attosecond
  spectroscopy in condensed matter,'' \emph{Nature}, vol. 449, no. 7165, p.
  1029, 2007. [Online]. Available: \url{http://dx.doi.org/10.1038/nature06229}
\BIBentrySTDinterwordspacing

\bibitem{neppl2012}
\BIBentryALTinterwordspacing
S.~Neppl, R.~Ernstorfer, E.~M. Bothschafter, A.~L. Cavalieri, D.~Menzel, J.~V.
  Barth, F.~Krausz, R.~Kienberger, and P.~Feulner, ``Attosecond time-resolved
  photoemission from core and valence states of magnesium,'' \emph{Phys. Rev.
  Lett.}, vol. 109, no.~8, pp. 087\,401--, Aug. 2012. [Online]. Available:
  \url{http://link.aps.org/doi/10.1103/PhysRevLett.109.087401}
\BIBentrySTDinterwordspacing

\bibitem{neppl2015}
\BIBentryALTinterwordspacing
S.~Neppl, R.~Ernstorfer, A.~L. Cavalieri, C.~Lemell, G.~Wachter, E.~Magerl,
  E.~M. Bothschafter, M.~Jobst, M.~Hofstetter, U.~Kleineberg, J.~V. Barth,
  D.~Menzel, J.~Burgdorfer, P.~Feulner, F.~Krausz, and R.~Kienberger, ``Direct
  observation of electron propagation and dielectric screening on the atomic
  length scale,'' \emph{Nature}, vol. 517, no. 7534, pp. 342--346, Jan. 2015.
  [Online]. Available: \url{http://dx.doi.org/10.1038/nature14094}
\BIBentrySTDinterwordspacing

\bibitem{locher2015}
\BIBentryALTinterwordspacing
R.~Locher, L.~Castiglioni, M.~Lucchini, M.~Greif, L.~Gallmann, J.~Osterwalder,
  M.~Hengsberger, and U.~Keller, ``Energy-dependent photoemission delays from
  noble metal surfaces by attosecond interferometry,'' \emph{Optica}, vol.~2,
  no.~5, pp. 405--410, 2015. [Online]. Available:
  \url{http://www.osapublishing.org/optica/abstract.cfm?URI=optica-2-5-405}
\BIBentrySTDinterwordspacing

\bibitem{lucchini2015}
\BIBentryALTinterwordspacing
M.~Lucchini, L.~Castiglioni, L.~Kasmi, P.~Kliuiev, A.~Ludwig, M.~Greif,
  J.~Osterwalder, M.~Hengsberger, L.~Gallmann, and U.~Keller, ``Light-matter
  interaction at surfaces in the spatiotemporal limit of macroscopic models,''
  \emph{Phys. Rev. Lett.}, vol. 115, p. 137401, Sep 2015. [Online]. Available:
  \url{https://link.aps.org/doi/10.1103/PhysRevLett.115.137401}
\BIBentrySTDinterwordspacing

\bibitem{seiffert2017}
\BIBentryALTinterwordspacing
L.~Seiffert, Q.~Liu, S.~Zherebtsov, A.~Trabattoni, P.~Rupp, M.~C. Castrovilli,
  M.~Galli, F.~Suszmann, K.~Wintersperger, J.~Stierle, G.~Sansone, L.~Poletto,
  F.~Frassetto, I.~Halfpap, V.~Mondes, C.~Graf, E.~Ruhl, F.~Krausz, M.~Nisoli,
  T.~Fennel, F.~Calegari, and M.~F. Kling, ``Attosecond chronoscopy of electron
  scattering in dielectric nanoparticles,'' \emph{Nat Phys}, vol.~13, pp. 766
  -- 770, 2017. [Online]. Available: \url{http://dx.doi.org/10.1038/nphys4129}
\BIBentrySTDinterwordspacing

\bibitem{PhysRevA.97.063415}
\BIBentryALTinterwordspacing
D.~Rattenbacher, I.~Jordan, A.~Schild, and H.~J. W\"orner, ``Nonlocal
  mechanisms of attosecond interferometry and implications for condensed-phase
  experiments,'' \emph{Phys. Rev. A}, vol.~97, p. 063415, Jun 2018. [Online].
  Available: \url{https://link.aps.org/doi/10.1103/PhysRevA.97.063415}
\BIBentrySTDinterwordspacing

\bibitem{McCurdy_2004}
\BIBentryALTinterwordspacing
C.~W. McCurdy, M.~Baertschy, and T.~N. Rescigno, ``Solving the three-body
  coulomb breakup problem using exterior complex scaling,'' \emph{Journal of
  Physics B: Atomic, Molecular and Optical Physics}, vol.~37, no.~17, pp.
  R137--R187, aug 2004. [Online]. Available:
  \url{https://doi.org/10.1088/0953-4075/37/17/R01}
\BIBentrySTDinterwordspacing

\bibitem{PhysRevA.62.032706}
\BIBentryALTinterwordspacing
T.~N. Rescigno and C.~W. McCurdy, ``Numerical grid methods for
  quantum-mechanical scattering problems,'' \emph{Phys. Rev. A}, vol.~62, p.
  032706, Aug 2000. [Online]. Available:
  \url{https://link.aps.org/doi/10.1103/PhysRevA.62.032706}
\BIBentrySTDinterwordspacing

\bibitem{PhysRevA.76.043420}
\BIBentryALTinterwordspacing
A.~Palacios, C.~W. McCurdy, and T.~N. Rescigno, ``Extracting amplitudes for
  single and double ionization from a time-dependent wave packet,'' \emph{Phys.
  Rev. A}, vol.~76, p. 043420, Oct 2007. [Online]. Available:
  \url{https://link.aps.org/doi/10.1103/PhysRevA.76.043420}
\BIBentrySTDinterwordspacing

\bibitem{scrinzi2012}
\BIBentryALTinterwordspacing
A.~Scrinzi, ``t-{SURFF}: fully differential two-electron photo-emission
  spectra,'' \emph{New Journal of Physics}, vol.~14, no.~8, p. 085008, aug
  2012. [Online]. Available:
  \url{https://doi.org/10.1088/1367-2630/14/8/085008}
\BIBentrySTDinterwordspacing

\bibitem{kroll1973}
\BIBentryALTinterwordspacing
N.~M. Kroll and K.~M. Watson, ``Charged-particle scattering in the presence of
  a strong electromagnetic wave,'' \emph{Phys. Rev. A}, vol.~8, pp. 804--809,
  Aug 1973. [Online]. Available:
  \url{https://link.aps.org/doi/10.1103/PhysRevA.8.804}
\BIBentrySTDinterwordspacing

\bibitem{faisal1973}
F.~H. Faisal, ``Multiple absorption of laser photons by atoms,'' \emph{J. Phys.
  B}, vol.~6, no.~4, p. L89, 1973.

\bibitem{madsen2005}
\BIBentryALTinterwordspacing
L.~B. Madsen, ``Strong-field approximation in laser-assisted dynamics,''
  \emph{Am. J. Phys.}, vol.~73, no.~1, pp. 57--62, 2005. [Online]. Available:
  \url{http://dx.doi.org/10.1119/1.1796791}
\BIBentrySTDinterwordspacing

\bibitem{galan2013}
\BIBentryALTinterwordspacing
A.~J. Gal\'an, L.~Argenti, and F.~Mart\'in, ``The soft-photon approximation in
  infrared-laser-assisted atomic ionization by extreme-ultraviolet
  attosecond-pulse trains,'' \emph{New Journal of Physics}, vol.~15, no.~11, p.
  113009, 2013. [Online]. Available:
  \url{http://stacks.iop.org/1367-2630/15/i=11/a=113009}
\BIBentrySTDinterwordspacing

\bibitem{PhysRevA.80.011201}
\BIBentryALTinterwordspacing
M.~A. McCune, M.~E. Madjet, and H.~S. Chakraborty, ``Reflective and collateral
  photoionization of an atom inside a fullerene: Confinement geometry from
  reciprocal spectra,'' \emph{Phys. Rev. A}, vol.~80, p. 011201, Jul 2009.
  [Online]. Available:
  \url{https://link.aps.org/doi/10.1103/PhysRevA.80.011201}
\BIBentrySTDinterwordspacing

\end{thebibliography}
\bibliographystyle{IEEEtran}

\end{document}